\documentclass[12pt]{article}
\def\be{\begin{equation}}
\def\ee{\end{equation}}
\def\ber{\begin{eqnarray}}
\def\eer{\end{eqnarray}}
\usepackage{amsfonts,amssymb}
\begin{document}
\vspace*{1cm}
\begin{center}
{\Large \bf Strange Mesons  in Kaluza-Klein Picture}\\

\vspace{4mm}

{\large A.A. Arkhipov\\
{\it State Research Center ``Institute for High Energy Physics" \\
 142280 Protvino, Moscow Region, Russia}}\\
\end{center}

\vspace{4mm}
\begin{abstract}
{We have performed an analysis of experimental data on mass
spectrum of the resonance states containing strange mesons and
compared them with the calculated values provided by Kaluza-Klein
scenario. In this note we present the results of this analysis.}
\end{abstract}

\section*{}

In our previous papers \cite{1,2} we have presented the arguments
in favour of that the Kaluza-Klein picture of the world has been
observed in the experiments at very low energies where the
nucleon-nucleon dynamics has been studied. In particular we have
found that geniusly simple formula for KK excitations provided by
Kaluza-Klein approach gives an excellent description for the mass
spectrum of two-nucleon system. In articles \cite{3,4} we have
presented additional arguments in favour of Kaluza and Klein
picture of the world. In fact, we have shown that simple formula
provided by Kaluza-Klein approach with the fundamental scale early
calculated \cite{1} gives an excellent description for the mass
spectrum of two-pion and three-pion systems. Now, taking this
line, we have performed an analysis of experimental data on mass
spectrum of the resonance states containing strange mesons and
compared them with the calculated values provided by Kaluza-Klein
scenario. In this note we present the results of this analysis.

Let us start with the study of two-kaon system. As in the previous
cases we build the Kaluza-Klein tower of KK excitations for
two-kaon system by the formula \be M_n^{K^1 K^2} =
\sqrt{m_{K^1}^2+\frac{n^2}{R^2}} +
\sqrt{m_{K^2}^2+\frac{n^2}{R^2}},\quad
(n=1,2,3,\ldots),\label{KK2K} \ee where $K^i(i=0,+,-)=K^0,K^+,K^-$
and $R$ is the same fundamental scale calculated early from the
analysis of nucleon-nucleon dynamics at low energies \cite{1,2}
\be \frac{1}{R} = 41.481\,\mbox{MeV}\quad \mbox{or}\quad
R=24.1\,\mbox{GeV}^{-1}=4.75\,10^{-13}\mbox{cm}.\label{scale} \ee
Kaluza-Klein tower such built is shown in Table 1 where the
comparison with experimentally observed mass spectrum of
$\phi$-mesons is also presented.

Throughout we have used Review of Particle Physics \cite{5} where the
experimental data on mass spectrum of the resonance states have been
extracted from. Some known experimental information is collected in
separate tables: Table 2 -- Table 7. We see from Tables~1--7 that
there is a quite remarkable correspondence of the calculated KK
excitations for two-kaon system with the experimentally observed mass
spectrum of the $\phi$-mesons. However, there are many empty cells in
Table 1 where we have not found the corresponding experimental data.
Mabe such data exist but we don't know these data. Anyway, we would
gratefully thank for any experimental information in this respect.

We also built the Kaluza-Klein tower of KK excitations for the
$K\pi$-system by the formula \be M_n^{K\pi} =
\sqrt{m_K^2+\frac{n^2}{R^2}} + \sqrt{m_\pi^2+\frac{n^2}{R^2}},\quad
(n=1,2,3,\ldots),\label{Kpi} \ee which is shown in Table 8 where the
comparison with experimentally observed mass spectrum is also
presented. Some known experimental information concerning the
experimentally observed resonance states in $K\pi$-system is
collected in separate tables: Table 9 -- Table 16. Again we see from
Tables 8--16 that there is a quite remarkable correspondence of the
calculated KK excitations for $K\pi$-system with the experimentally
observed mass spectrum of the resonance states in $K\pi$-system.
Here, there are empty cells in Table 8 as well, where we have not
found the corresponding experimental data.

Calculating the Kaluza-Klein tower of KK excitations for the
$K2\pi$-system by the formula

\be M_n^{K2\pi} = \sqrt{m_K^2+\frac{n^2}{R^2}} +
2\sqrt{m_\pi^2+\frac{n^2}{R^2}},\quad (n=1,2,3,\ldots),\label{K2pi}
\ee is shown in Table 17 where the comparison with experimentally
observed mass spectrum has been presented too. Some known
experimental information concerning the experimentally observed
resonance states in $K\pi\pi$-system is collected in separate tables:
Table 18 -- Table 20.

As in previous history we see from Tables 17--20 a quite remarkable
correspondence of the calculated KK excitations for $K\pi\pi$-system
with the masses of the resonance states in $K\pi\pi$-system where
such resonance states are experimentally observed. Many empty cells
in Table 17 indicate a wide field in experimental study of
$K\pi\pi$-system.

Here we also present the results calculating the Kaluza-Klein
tower of KK excitations for the $K\rho$-system by the formula

\begin{equation}
M_n^{K\rho} = \sqrt{m_K^2+\frac{n^2}{R^2}} +
\sqrt{m_\rho^2+\frac{n^2}{R^2}},\quad
(n=1,2,3,\ldots).\label{Krho}
\end{equation}
These results are shown in Table 21 together with existing
experimental data. As usual more detailed experimental information
concerning the resonance states in $K\rho$-system is collected in
separate tables: Table 22 -- Table 28.

Finally, we built the Kaluza-Klein tower of KK excitations for the
$K\eta$-system by the formula
\begin{equation}
M_n^{K\eta} = \sqrt{m_K^2+\frac{n^2}{R^2}} +
\sqrt{m_\eta^2+\frac{n^2}{R^2}},\quad
(n=1,2,3,\ldots).\label{Keta}
\end{equation}
This is shown in Table 29. Here we have found only one experimental
point (see Table 30) corresponding to $M_{17}^{K\eta}$-- storey. That
is why, it needs the further experimental study of $K\eta$-system.

From analysis performed we could once more emphasize a remarkable
fact mentioned in our previous papers: the resonances with the
different quantum numbers may occupy one and the same storey in KK
tower.\footnote{Similar (``mass-band") structures were empirically
observed early in Ref. \cite{6}. I thank D.~Akers for drawing my
attention to this article.} This is a peculiarity of the
systematics provided by Kaluza-Klein picture.

We have also established here a new remarkable feature of
Kaluza-Klein picture which manifests itself in the existence of
intersecting mass terms corresponding to the states of the
hadronic systems with the different mesons content. We enumerate
such intersecting mass terms below.

\centerline{\bf\large{Intersecting mass terms}}
\begin{itemize}
\item
$M_3^{2K}(1018.24-1025.99)\,\bigcap\, M_9^{K\pi}(1015.93-1020.70)
\neq\varnothing $
\item
$M_5^{2K}(1070.95-1078.32)\,\bigcap\, M_3^{K\eta}(1070.39-1074.26)
\neq\varnothing $
\item
$M_7^{2K}(1145.78-1152.37)\,\bigcap\,
M_{11}^{K\pi}(1148.09-1152.35) \neq\varnothing $
\item
$M_8^{2K}(1189.69-1196.33)\,\bigcap\, M_7^{K\eta}(1192.30-1195.74)
\neq\varnothing $
\item
$M_9^{2K}(1237.89-1244.27)\,\bigcap\, M_8^{K\eta}(1234.89-1238.21)
\neq\varnothing $
\item
$M_{10}^{2K}(1289.63-1295.76)\,\bigcap\,
M_{13}^{K\pi}(1287.00-1290.83)\neq\varnothing $
\item
$M_{10}^{2K}(1289.63-1295.76)\,\bigcap\,
M_{3}^{K\rho}(1288.42-1292.29) \neq\varnothing $
\item
$M_{17}^{2K}(1721.62-1726.22)\,\bigcap\,
M_{14}^{K\rho}(1726.10-1728.70) \neq\varnothing $
\item
$M_{25}^{2K}(2297.08-2300.53)\,\bigcap\,
M_{17}^{K2\pi}(22976.78-2300.83)\neq\varnothing $
\item
$M_{25}^{2K}(2297.08-2300.53)\,\bigcap\,
M_{23}^{K\rho}(2299.82-2301.66) \neq\varnothing $
\item
$M_5^{K\pi}(782.93-789.16)\,\bigcap\, M_1^{K2\pi}(777.83-790.61)
\neq\varnothing $
\item
$M_8^{K\pi}(953.10-958.17)\,\bigcap\, M_4^{K2\pi}(948.60-958.24)
\neq\varnothing $
\item
$M_{12}^{K\pi}(1216.82-1220.86)\,\bigcap\,
M_7^{K2\pi}(1213.15-1220.53) \neq\varnothing $
\item
$M_{13}^{K\pi}(1287.00-1290.83)\,\bigcap\,
M_3^{K\rho}(1288.42-1292.29) \neq\varnothing $
\item
$M_{14}^{K\pi}(1358.43-1362.08)\,\bigcap\,
M_6^{K\rho}(1361.43-1365.00) \neq\varnothing $
\item
$M_{15}^{K\pi}(1430.97-1434.44)\,\bigcap\,
M_8^{K\rho}(1432.67-1435.99) \neq\varnothing $
\item
$M_{22}^{K\pi}(1960.08-1962.67)\,\bigcap\,
M_{20}^{K\eta}(1959.29-1961.34) \neq\varnothing $
\item
$M_{29}^{K\pi}(2510.82-2512.86)\,\bigcap\,
M_{26}^{K\rho}(2510.90-2512.57) \neq\varnothing $
\item
$M_{6}^{K2\pi}(1119.13-1127.14)\,\bigcap\,
M_{5}^{K\eta}(1120.76-1124.44) \neq\varnothing $
\item
$M_{8}^{K2\pi}(1311.35-1318.18)\,\bigcap\,
M_{4}^{K\rho}(1307.81-1311.59) \neq\varnothing $
\item
$M_{10}^{K2\pi}(1517.25-1523.21)\,\bigcap\,
M_{10}^{K\rho}(1518.82-1521.89) \neq\varnothing $
\item
$M_{11}^{K2\pi}(1623.93-1629.51)\,\bigcap\,
M_{15}^{K\eta}(1622.94-1625.43) \neq\varnothing $
\item
$M_{13}^{K2\pi}(1842.89-1847.86)\,\bigcap\,
M_{16}^{K\rho}(1843.20-1845.59) \neq\varnothing $
\item
$M_{17}^{K2\pi}(2296.78-2300.83)\,\bigcap\,
M_{23}^{K\rho}(2299.82-2301.66) \neq\varnothing $
\end{itemize}

Of course, it would be very desirable to state new experiments to
search new states corresponding to the empty cells in Tables
1,8,17,21,29. These Tables may serve as a guide for the
physicists--experimenters. We believe that this is a quite promising
subject of the investigations in particle and nuclear physics.


\newpage

\begin{center}
Table 1: Kaluza-Klein tower of KK excitations in $KK$-system and
experimental data.

\vspace{5mm}
{\large
\begin{tabular}{|c|c|c|c|c|}\hline
 n & $ M_n^{2K^0}$MeV & $ M_n^{K^0 K^\pm}$MeV & $
 M_n^{2K^\pm}$MeV & $ M_{exp}^{2K}$\,MeV  \\
 \hline
1  & 998.80  & 994.81  & 990.83  &   \\
2  & 1009.08 & 1005.14 & 1001.20 &   \\
3  & 1025.99 & 1022.11 & 1018.24 & 1019.417$\pm$0.014 \\
4  & 1049.21 & 1045.42 & 1041.63 &    \\
5  & 1078.32 & 1074.64 & 1070.95 &   \\
6  & 1112.87 & 1109.30 & 1105.73 &  \\
7  & 1152.37 & 1148.93 & 1145.48 &    \\
8  & 1196.33 & 1193.01 & 1189.69 &    \\
9  & 1244.27 & 1241.08 & 1237.89 &  \\
10 & 1295.76 & 1292.69 & 1289.63 &    \\
11 & 1350.38 & 1347.44 & 1344.50 & 1346 - i249 \\
12 & 1407.77 & 1404.96 & 1402.14 &  \\
13 & 1467.62 & 1464.91 & 1462.21 & 1463 $\pm$ 9  \\
14 & 1529.62 & 1527.02 & 1524.43 &    \\
15 & 1593.53 & 1591.04 & 1588.55 &   \\
16 & 1659.13 & 1656.74 & 1654.34 & 1655 $\pm$ 17 \\
17 & 1726.22 & 1723.92 & 1721.62 &  \\
18 & 1794.64 & 1792.43 & 1790.22 &  \\
19 & 1864.24 & 1862.11 & 1859.99 & 1864.1 $\pm$ 1.0  \\
20 & 1934.89 & 1932.85 & 1930.80 &  \\
21 & 2006.49 & 2004.52 & 2002.54 & 2006.7 $\pm$ 0.5 \\
22 & 2078.93 & 2077.03 & 2075.12 &   \\
23 & 2152.14 & 2150.30 & 2148.45 &  \\
24 & 2226.02 & 2224.24 & 2222.46 &   \\
25 & 2300.53 & 2298.81 & 2297.08 &   \\
26 & 2375.60 & 2373.93 & 2372.26 &   \\
27 & 2451.17 & 2449.56 & 2447.94 & \\
28 & 2527.21 & 2525.64 & 2524.08 & \\
29 & 2603.67 & 2602.15 & 2600.63 &   \\
30 & 2680.52 & 2679.04 & 2677.57 &  \\ \hline
\end{tabular}}
\end{center}

\newpage

\begin{center}
Table 2: $M_{3}^{2K}(1018-1023)$--Storey.

\vspace{5mm}
\begin{tabular}{|c|c|c|c|c|}\hline
$R(I^GJ^{PC})$ & $ M_R $\, MeV & $ \Gamma_R $\, MeV & Reaction &
Collab. \\ \hline $\phi(0^-1^{--})$ & 1019.417$\pm$0.014 & 4.458
$\pm$ 0.032 & AVERAGE & PDG 00 \\ \hline
\end{tabular}
\end{center}

\vspace{10mm}
\begin{center}
Table 3: $M_{11}^{2K}(1345-1350)$--Storey.

\vspace{5mm}
\begin{tabular}{|c|c|c|c|c|}\hline
$R(I^GJ^{PC})$ & $ M_R $\, MeV & $ \Gamma_R $\, MeV & Reaction &
Collab. \\ \hline $f_0(0^+0^{++})$ & 1346 - i249 & 2Im$M_R$  &
$\pi\pi \rightarrow \pi\pi,K{\bar K} $ & RVUE 95
\\ \hline
\end{tabular}
\end{center}

\vspace{10mm}
\begin{center}
Table 4: $M_{13}^{2K}(1462-1468)$--Storey.

\vspace{5mm}
\begin{tabular}{|c|c|c|c|c|}\hline
$R(I^GJ^{PC})$ & $ M_R $\, MeV & $ \Gamma_R $\, MeV & Reaction &
Collab. \\ \hline $f_0(0^+0^{++}$ & 1463$\pm$9 &
$118^{\,+\,138}_{\,-\,16} $ & $\pi^-p \rightarrow 2K_S^0n $ & MPS 82
\\ \hline
\end{tabular}
\end{center}

\vspace{10mm}
\begin{center}
Table 5: $M_{16}^{2K}(1654-1659)$--Storey.

\vspace{5mm}
\begin{tabular}{|c|c|c|c|r|}\hline
$R(I^GJ^{PC})$ & $ M_R $\, MeV & $ \Gamma_R $\, MeV & Reaction &
Collab.\ \ \ \\ \hline $\phi(0^-1^{--})$ & 1657 $\pm$ 27 & 146
$\pm$ 55 & $ e^+ e^-
\rightarrow K_S^0 K^{\pm}\pi^{\mp} $ & DM2 91 \\
 & 1655 $\pm$ 17 & 207 $\pm$ 45 & $ e^+ e^-
\rightarrow K^+ K^-$ & DM2 88 \\
 & 1681 $\pm$ 8 & 150 $\pm$ 50 & AVERAGE & PDG 00 \\ \hline
\end{tabular}
\end{center}

\vspace{10mm}
\begin{center}
Table 6: $M_{19}^{2K}(1860-1864)$--Storey.

\vspace{5mm}
\begin{tabular}{|c|c|c|c|r|}\hline
$R(I^GJ^{PC})$ & $ M_R $\, MeV & $ \Gamma_R $\, MeV & Reaction &
Collab.\ \  \\ \hline $\phi(0^-3^{--})$ & 1855 $\pm$ 10 & 64 $\pm$
31 & $
K^- p \rightarrow K^- K^+\Lambda $ & LASS 88 \\
 & $1870^{\,+\,30}_{\,-\,20}$ & $160^{\,+\,90}_{\,-\,50}$ & $ K^- p
\rightarrow K^- K^+\Lambda $ & OMEG 82 \\
 & 1850 $\pm$ 10 & $80^{\,+\,40}_{\,-\,30}$ & $ K^- p
\rightarrow K{\bar K}\Lambda $ & HBC 81 \\
 & 1854 $\pm$ 7 & $87^{\,+\,28}_{\,-\,23}$ & AVERAGE & PDG 00 \\
$D^0(\frac{1}{2}0^{-})$ & 1864.1 $\pm$ 1.0 &  & AVERAGE & PDG 00 \\
\hline
\end{tabular}
\end{center}

\vspace{10mm}
\begin{center}
Table 7: $M_{21}^{2K}(2003-2006)$--Storey.

\vspace{5mm}
\begin{tabular}{|c|c|c|c|r|}\hline
$R(IJ^{P})$ & $ M_R $\, MeV & $ \Gamma_R $\, MeV & Reaction &
Collab.\ \ \\ \hline $D^*(\frac{1}{2}1^{-})$ & 2006.7 $\pm$ 0.5 &
$<$ 2.1 & AVERAGE & PDG 00
\\ \hline
\end{tabular}
\end{center}

\newpage

\begin{center}
Table 8: Kaluza-Klein tower of KK excitations in $K\pi$-system and
experimental data.

\vspace{5mm} {\large
\begin{tabular}{|c|c|c|c|c|c|}\hline
 n & $ M_n^{K^0 \pi^0}$MeV & $ M_n^{K^0 \pi^{\pm}}$MeV & $
 M_n^{K^\pm \pi^0}$MeV & $ M_n^{K^\pm \pi^\pm}$MeV & $
 M_{exp}^{K \pi}$\,MeV  \\
 \hline
1  & 640.60  & 645.00  & 636.62  & 641.02  &  \\
2  & 662.97  & 666.91  & 659.03  & 662.97  &   \\
3  & 696.58  & 699.99  & 692.71  & 696.11  &  \\
4  & 738.50  & 741.42  & 734.71  & 737.63  &  \\
5  & 786.62  & 789.16  & 782.93  & 785.47  &  \\
6  & 839.57  & 841.79  & 836.00  & 838.22  &  \\
7  & 896.39  & 898.36  & 892.95  & 894.91  & $K^*(892)$ \\
8  & 956.42  & 958.17  & 953.10  & 954.85  &  \\
9  & 1019.12 & 1020.70 & 1015.93 & 1017.51 &  \\
10 & 1084.10 & 1085.54 & 1081.03 & 1082.48 &  \\
11 & 1151.03 & 1152.35 & 1148.09 & 1149.41 &  \\
12 & 1219.64 & 1220.86 & 1216.82 & 1218.04 &  \\
13 & 1289.70 & 1290.83 & 1287.00 & 1288.13 &  \\
14 & 1361.03 & 1362.08 & 1358.43 & 1359.48 &  \\
15 & 1433.45 & 1434.44 & 1430.97 & 1431.95 & $K_{0,2}^*(1430)$ \\
16 & 1506.85 & 1507.78 & 1504.46 & 1505.39 &  \\
17 & 1581.09 & 1581.97 & 1578.79 & 1579.67 & $K_2(1580)$ \\
18 & 1656.08 & 1656.91 & 1653.87 & 1654.70 & $K^*(1680)$ \\
19 & 1731.74 & 1732.53 & 1729.61 & 1730.40 & $K_3^*(1780)$ \\
20 & 1807.98 & 1808.73 & 1805.93 & 1806.68 &  \\
21 & 1884.75 & 1885.46 & 1882.77 & 1883.49 &  \\
22 & 1961.98 & 1962.67 & 1960.08 & 1960.76 & $K_0^*(1950)$ \\
23 & 2039.64 & 2040.29 & 2037.80 & 2038.45 & $K_4^*(2045)$ \\
24 & 2117.67 & 2118.30 & 2115.89 & 2116.52 &  \\
25 & 2196.04 & 2196.65 & 2194.32 & 2194.92 &  \\
26 & 2274.72 & 2275.30 & 2273.06 & 2273.64 &  \\
27 & 2353.68 & 2354.24 & 2352.07 & 2352.63 & $K_5^*(2380)$ \\
28 & 2432.90 & 2433.44 & 2431.33 & 2431.87 &  \\
29 & 2512.34 & 2512.86 & 2510.82 & 2511.34 &  \\
30 & 2592.00 & 2592.50 & 2590.52 & 2591.02 &  \\ \hline
\end{tabular}}
\end{center}

\newpage

\begin{center}
Table 9: $M_{7}^{K\pi}(893-898)$--Storey.

\vspace{3mm}
\begin{tabular}{|c|c|c|c|r|}\hline
$R(IJ^{P})$ & $ M_R $\, MeV & $ \Gamma_R $\, MeV & Reaction &
Collab.\ \ \\ \hline $K^*(\frac{1}{2}1^{-})$ & 896.1 $\pm$ 0.27 &
50.7 $\pm$ 0.6 & AVERAGE & PDG 00 \\ \hline
\end{tabular}
\end{center}

\vspace{3mm}
\begin{center}
Table 10: $M_{15}^{K\pi}(1431-1434)$--Storey.

\vspace{3mm}
\begin{tabular}{|c|c|c|c|r|}\hline
$R(IJ^{P})$ & $ M_R $\, MeV & $ \Gamma_R $\, MeV & Reaction &
Collab.\ \ \\ \hline $K_0^*(\frac{1}{2}0^{+})$ & 1436 $\pm$ 8 &
196 $\pm$ 45 & $pp
\rightarrow p_f p_s K^+ K^- \pi^+\pi^- $ & OMEG 98 \\
 & 1415 $\pm$ 25 & 330 $\pm$ 50 & $K^- p
\rightarrow K^- \pi^+ n $ & RVUE 97 \\
 & $\sim$ 1430 & $\sim$ 200 & $K^- p
\rightarrow {\bar K}^0 \pi^- p $ & HBC 84 \\
$K_2^*(\frac{1}{2}2^{+})$ & 1431.2$\pm$1.8$\pm$0.7 &
116.5$\pm$3.6$\pm$1.7 & $K^- p
\rightarrow  K^- \pi^+ n $ & LASS 88 \\
 & 1432.4 $\pm$ 1.3 & 109 $\pm$ 5$\pm$1.7 & AVERAGE & PDG 00 \\
\hline
\end{tabular}
\end{center}

\vspace{3mm}
\begin{center}
Table 11: $M_{17}^{K\pi}(1578-1582)$--Storey.

\vspace{3mm}
\begin{tabular}{|c|c|c|c|r|}\hline
$R(IJ^{P})$ & $ M_R $\, MeV & $ \Gamma_R $\, MeV & Reaction &
Collab.\ \ \\ \hline $K_2(\frac{1}{2}2^{-})$ & $\sim$  1580 &
$\sim$ 110 & $K^{-}p$ & AACH3 79 \\ \hline
\end{tabular}
\end{center}

\vspace{3mm}
\begin{center}
Table 12: $M_{18}^{K\pi}(1654-1657)$--Storey.

\vspace{3mm}
\begin{tabular}{|c|c|c|c|r|}\hline
$R(IJ^{P})$ & $ M_R $\, MeV & $ \Gamma_R $\, MeV & Reaction &
Collab.\ \ \\ \hline $K^*(\frac{1}{2}1^{-})$ & 1677$\pm$10$\pm$32
& 205$\pm$16$\pm$34 &
$K^{-}p \rightarrow  K^- \pi^+ n$ & LASS 88 \\
 & $\sim$  1650 & 250-300 & $K^{\pm}p \rightarrow  K^{\pm}\pi^{\pm}n$
 & ASPK 78 \\
\hline
\end{tabular}
\end{center}

\vspace{3mm}
\begin{center}
Table 13: $M_{19}^{K\pi}(1730-1733)$--Storey.

\vspace{3mm}
\begin{tabular}{|c|c|c|c|r|}\hline
$R(IJ^{P})$ & $ M_R $\, MeV & $ \Gamma_R $\, MeV & Reaction &
Collab.\ \ \\ \hline $K_3^*(\frac{1}{2}3^{-})$ &
1740$\pm$14$\pm$15 & 171$\pm$42$\pm$20 &
$K^{-}p \rightarrow  {\bar K}^0 \pi^+\pi^- n$ & LASS 87 \\
 & 1720$\pm$10$\pm$15 & 187$\pm$31$\pm$20 & $K^{-}p \rightarrow
 {\bar K}^0 \pi^- p$ & LASS 89 \\ \hline
\end{tabular}
\end{center}

\vspace{3mm}
\begin{center}
Table 14: $M_{22}^{K\pi}(1960-1963)$--Storey.

\vspace{3mm}
\begin{tabular}{|c|c|c|c|r|}\hline
$R(IJ^{P})$ & $ M_R $\, MeV & $ \Gamma_R $\, MeV & Reaction &
Collab.\ \ \\ \hline $K_0^*(\frac{1}{2}0^{+})$ &
1945$\pm$10$\pm$20 & 201$\pm$34$\pm$79 & $K^{-}p \rightarrow K^-
\pi^+ n$ & LASS 88 \\ \hline
\end{tabular}
\end{center}

\vspace{3mm}
\begin{center}
Table 15: $M_{23}^{K\pi}(2038-2040)$--Storey.

\vspace{3mm}
\begin{tabular}{|c|c|c|c|r|}\hline
$R(IJ^{P})$ & $ M_R $\, MeV & $ \Gamma_R $\, MeV & Reaction &
Collab.\ \ \\ \hline $K_4^*(\frac{1}{2}4^{+})$ & 2039 $\pm$ 10 &
189 $\pm$ 35 & $K^{+}p
\rightarrow K_S^0 \pi^{\pm}p$ & SPEC 82 \\
 & 2062$\pm$14$\pm$13 & 221$\pm$48$\pm$27 & $K^{-}p \rightarrow K^-
 \pi^+ n$ & LASS 86 \\
 & 2045 $\pm$ 9 & 198 $\pm$ 30 & AVERAGE & PDG 00 \\
 \hline
\end{tabular}
\end{center}

\vspace{3mm}
\begin{center}
Table 16: $M_{27}^{K\pi}(2352-2354)$--Storey.

\vspace{3mm}
\begin{tabular}{|c|c|c|c|r|}\hline
$R(IJ^{P})$ & $ M_R $\, MeV & $ \Gamma_R $\, MeV & Reaction &
Collab.\ \ \\ \hline $K_5^*(\frac{1}{2}5^{-})$ &
2382$\pm$14$\pm$19 & 178$\pm$37$\pm$32 & $K^{-}p \rightarrow K^-
\pi^+ n$ & LASS 86 \\ \hline
\end{tabular}
\end{center}

\newpage

\begin{center}
Table 17: Kaluza-Klein tower of KK excitations in $K2\pi$-system and
experimental data.

\vspace{5mm} {\large
\begin{tabular}{|c|c|c|c|c|c|}\hline
 n & $ M_n^{K^0 2\pi^0}$MeV & $ M_n^{K^0 2\pi^{\pm}}$MeV &
 $M_n^{K^\pm 2\pi^0}$MeV & $ M_n^{K^\pm 2\pi^\pm}$MeV & $
 M_{exp}^{K 2\pi}$\,MeV  \\
 \hline
1  & 781.81  & 790.61  & 777.83  & 786.62  &  \\
2  & 821.41  & 829.27  & 817.47  & 825.33  &  \\
3  & 880.17  & 886.98  & 876.30  & 883.10  &  \\
4  & 952.39  & 958.24  & 948.60  & 954.45  &  \\
5  & 1034.08 & 1039.15 & 1030.39 & 1035.46 &  \\
6  & 1122.70 & 1127.14 & 1119.13 & 1123.57 &  \\
7  & 1216.60 & 1220.53 & 1213.15 & 1217.08 &  \\
8  & 1314.67 & 1318.18 & 1311.35 & 1314.86 &  \\
9  & 1416.10 & 1419.27 & 1412.91 & 1416.08 &  \\
10 & 1520.32 & 1523.21 & 1517.25 & 1520.14 &  \\
11 & 1626.87 & 1629.51 & 1623.93 & 1626.57 & 1629 $\pm$ 7 \\
12 & 1735.39 & 1737.83 & 1732.57 & 1735.01 & 1730 $\pm$ 20 \\
13 & 1845.59 & 1847.86 & 1842.89 & 1845.15 & $\sim$ 1840 \\
14 & 1957.24 & 1959.36 & 1954.65 & 1956.76 &  \\
15 & 2070.14 & 2072.12 & 2067.66 & 2069.63 &  \\
16 & 2184.13 & 2186.00 & 2181.74 & 2183.60 &  \\
17 & 2299.07 & 2300.83 & 2296.78 & 2298.53 &  \\
18 & 2414.85 & 2416.51 & 2412.64 & 2414.30 &  \\
19 & 2531.36 & 2532.93 & 2529.23 & 2530.81 &  \\
20 & 2648.51 & 2650.01 & 2646.46 & 2647.96 &  \\
21 & 2766.25 & 2767.68 & 2764.27 & 2765.70 &  \\
22 & 2884.50 & 2885.86 & 2882.59 & 2883.96 &  \\
23 & 3003.21 & 3004.51 & 3001.36 & 3002.67 &  \\
24 & 3122.33 & 3123.58 & 3120.55 & 3121.80 &  \\
25 & 3241.82 & 3243.03 & 3240.10 & 3241.30 &  \\
26 & 3361.65 & 3362.81 & 3359.98 & 3361.14 &  \\
27 & 3481.78 & 3482.90 & 3480.16 & 3481.28 &  \\
28 & 3602.19 & 3603.27 & 3600.62 & 3601.70 &  \\
29 & 3722.85 & 3723.89 & 3721.32 & 3722.37 &  \\
30 & 3843.73 & 3844.74 & 3842.25 & 3843.26 &  \\ \hline
\end{tabular}}
\end{center}

\newpage

\vspace*{50mm}
\begin{center}
Table 18: $M_{11}^{K2\pi}(1624-1630)$--Storey.

\vspace{5mm}
\begin{tabular}{|c|c|c|c|r|}\hline
$R(IJ^{P})$ & $ M_R $\, MeV & $ \Gamma_R $\, MeV & Reaction &
Collab.\ \ \\ \hline $K(\frac{1}{2}?^{?})$ & 1629 $\pm$ 7 &
$16^{\,+\,19}_{\,-\,16}$ & $\pi^{-}p \rightarrow K_S^0 \pi^+\pi^-$
& BC 98 \\ \hline
\end{tabular}
\end{center}

\vspace{10mm}

\begin{center}
Table 19: $M_{12}^{K2\pi}(1733-1738)$--Storey.

\vspace{5mm}
\begin{tabular}{|c|c|c|c|r|}\hline
$R(IJ^{P})$ & $ M_R $\, MeV & $ \Gamma_R $\, MeV & Reaction &
Collab.\ \ \\ \hline $K_2(\frac{1}{2}2^{-})$ & 1730 $\pm$ 20 & 210
$\pm$ 30 & $K^{+}d$ & DBC 72 \\
 & 1740  & 130 & $K^{-}d \rightarrow {\bar K}2\pi d$ & DBC 71 \\
 & 1745 $\pm$ 20 & 100 $\pm$ 50 & $K^- p$ & HBC 70 \\
 & 1765 $\pm$ 40 & 90 $\pm$ 70 & $K^{+}p \rightarrow  K2\pi N$ & HBC 71 \\
 \hline
\end{tabular}
\end{center}

\vspace{10mm}
\begin{center}
Table 20: $M_{13}^{K2\pi}(1843-1848)$--Storey.

\vspace{5mm}
\begin{tabular}{|c|c|c|c|r|}\hline
$R(IJ^{P})$ & $ M_R $\, MeV & $ \Gamma_R $\, MeV & Reaction &
Collab.\ \ \\ \hline $K_1(\frac{1}{2}1^{+})$ & $\sim$ 1840 &
$\sim$ 250
 & $K^{-}p \rightarrow 3Kp$ & OMEG 83 \\
 & $\sim$ 1800 & $\sim$ 250 & $K^{-}p \rightarrow K^- 2\pi p$ & CNTR 81 \\
$K_2(\frac{1}{2}2^{-})$ & $\sim$ 1840 & $\sim$ 230 & $K^{-}p
\rightarrow K^- 2\pi p$ & CNTR 81 \\ \hline
\end{tabular}
\end{center}

\newpage

\begin{center}
Table 21: Kaluza-Klein tower of KK excitations in $K\rho$-system and
experimental data.

\vspace{5mm} {\large
\begin{tabular}{|c|c|c|c|}\hline
 n & $ M_n^{K^0\rho}$\,MeV & $ M_n^{K^\pm\rho}$\,MeV & $
 M_{exp}^{K\rho}$\,MeV   \\
 \hline
1  & 1269.82 & 1265.83 &  \\
2  & 1278.30 & 1274.36 & 1273 $\pm$ 7 \\
3  & 1292.29 & 1288.42 &  \\
4  & 1311.59 & 1307.81 &  \\
5  & 1335.93 & 1332.24 &  \\
6  & 1365.00 & 1361.43 &  \\
7  & 1398.46 & 1395.01 & 1402 $\pm$ 7 \\
8  & 1435.99 & 1432.67 & 1414 $\pm$ 15 \\
9  & 1477.24 & 1474.05 & $\sim$ 1460 \\
10 & 1521.89 & 1518.82 &  \\
11 & 1569.63 & 1566.69 &  \\
12 & 1620.19 & 1617.37 &  \\
13 & 1673.29 & 1670.58 &  \\
14 & 1728.70 & 1726.10 & 1717 $\pm$ 27 \\
15 & 1786.20 & 1783.71 & 1776 $\pm$ 7 \\
16 & 1845.59 & 1843.20 &  \\
17 & 1906.71 & 1904.41 &  \\
18 & 1969.39 & 1967.18 & 1973$\pm$8$\pm$25 \\
19 & 2033.48 & 2031.35 &  \\
20 & 2098.86 & 2096.81 &  \\
21 & 2165.42 & 2163.44 &  \\
22 & 2233.05 & 2231.14 &  \\
23 & 2301.66 & 2299.82 &  \\
24 & 2371.16 & 2369.38 &  \\
25 & 2441.49 & 2439.76 &  \\
26 & 2512.57 & 2510.90 &  \\
27 & 2584.34 & 2582.72 &  \\
28 & 2656.75 & 2655.18 &  \\
29 & 2729.75 & 2728.22 &  \\
30 & 2803.29 & 2801.81 &  \\ \hline
\end{tabular}}
\end{center}

\newpage

\begin{center}
Table 22: $M_{2}^{K\rho}(1274-1278)$--Storey.

\vspace{5mm}
\begin{tabular}{|c|c|c|c|r|}\hline
$R(IJ^{P})$ & $ M_R $\, MeV & $ \Gamma_R $\, MeV & Reaction &
Collab.\ \ \\ \hline $K_1(\frac{1}{2}1^{+})$ & 1275 $\pm$ 10 & 75
$\pm$ 15 & $K^{-}p \rightarrow \Xi(K\pi\pi)^{+}$ & HBC 78 \\
 & 1273 $\pm$ 7  & 90 $\pm$ 20 & AVERAGE & PDG 00 \\ \hline
\end{tabular}
\end{center}

\vspace{5mm}
\begin{center}
Table 23: $M_{7}^{K\rho}(1395-1398)$--Storey.

\vspace{5mm}
\begin{tabular}{|c|c|c|c|r|}\hline
$R(IJ^{P})$ & $ M_R $\, MeV & $ \Gamma_R $\, MeV & Reaction &
Collab.\ \ \\ \hline $K_1(\frac{1}{2}1^{+})$ & 1392 $\pm$ 18 & 276
$\pm$ 65 & $K^{-}p \rightarrow K^0_S\pi^{+}\pi^{-}n$ & HBC 82 \\
 & 1402 $\pm$ 7  & 174 $\pm$ 13 & AVERAGE & PDG 00 \\ \hline
\end{tabular}
\end{center}

\vspace{5mm}
\begin{center}
Table 24: $M_{8}^{K\rho}(1433-1436)$--Storey.

\vspace{5mm}
\begin{tabular}{|c|c|c|c|r|}\hline
$R(IJ^{P})$ & $ M_R $\, MeV & $ \Gamma_R $\, MeV & Reaction &
Collab.\ \ \\ \hline $K^*(\frac{1}{2}1^{-})$ & 1420$\pm$7$\pm$10 &
240$\pm$18$\pm$12 & $K^{-}p \rightarrow {\bar K}^0\pi^{+}\pi^{-}n$ & LASS 87 \\
 & 1414 $\pm$ 15  & 232 $\pm$ 21 & AVERAGE & PDG 00 \\ \hline
\end{tabular}
\end{center}

\vspace{5mm}
\begin{center}
Table 25: $M_{9}^{K\rho}(1474-1477)$--Storey.

\vspace{5mm}
\begin{tabular}{|c|c|c|c|r|}\hline
$R(IJ^{P})$ & $ M_R $\, MeV & $ \Gamma_R $\, MeV & Reaction &
Collab.\ \ \\ \hline $K(\frac{1}{2}0^{-})$ & $\sim$ 1460 & $\sim$
260 & $K^{-}p \rightarrow K^{-}2\pi p$ & CNTR 81 \\ \hline
\end{tabular}
\end{center}

\vspace{5mm}
\begin{center}
Table 26: $M_{14}^{K\rho}(1726-1729)$--Storey.

\vspace{5mm}
\begin{tabular}{|c|c|c|c|r|}\hline
$R(IJ^{P})$ & $ M_R $\, MeV & $ \Gamma_R $\, MeV & Reaction &
Collab.\ \ \\ \hline $K^*(\frac{1}{2}1^{-})$ & 1735$\pm$10$\pm$20
& 423$\pm$18$\pm$30 & $K^{-}p \rightarrow {\bar K}^0 \pi^{+}\pi^{-}n$ & LASS 87 \\
 & 1717 $\pm$ 27  & 322 $\pm$ 110 & AVERAGE & PDG 00 \\
\hline
\end{tabular}
\end{center}

\vspace{5mm}
\begin{center}
Table 27: $M_{15}^{K\rho}(1784-1786)$--Storey.

\vspace{5mm}
\begin{tabular}{|c|c|c|c|r|}\hline
$R(IJ^{P})$ & $ M_R $\, MeV & $ \Gamma_R $\, MeV & Reaction &
Collab.\ \ \\ \hline $K_3^*(\frac{1}{2}3^{-})$ & 1776 $\pm$ 7 &
159 $\pm$ 21 & AVERAGE & PDG 00 \\
\hline
\end{tabular}
\end{center}

\vspace{5mm}
\begin{center}
Table 28: $M_{18}^{K\rho}(1967-1969)$--Storey.

\vspace{5mm}
\begin{tabular}{|c|c|c|c|r|}\hline
$R(IJ^{P})$ & $ M_R $\, MeV & $ \Gamma_R $\, MeV & Reaction &
Collab.\ \ \\ \hline $K_2^*(\frac{1}{2}2^{+})$ & 1973$\pm$8$\pm$25
& 373$\pm$33$\pm$60 & AVERAGE & PDG 00 \\
\hline
\end{tabular}
\end{center}

\newpage

\begin{center}
Table 29: Kaluza-Klein tower of KK excitations in $K\eta$-system and
experimental data.

\vspace{5mm} {\large
\begin{tabular}{|c|c|c|c|}\hline
 n & $ M_n^{K^0\eta}$\,MeV & $ M_n^{K^\pm\eta}$\,MeV & $
 M_{exp}^{K\eta}$\,MeV   \\
 \hline
1  & 1048.27 & 1044.29 &  \\
2  & 1058.09 & 1054.15 &  \\
3  & 1074.26 & 1070.39 &  \\
4  & 1096.50 & 1092.71 &  \\
5  & 1124.44 & 1120.76 &  \\
6  & 1157.67 & 1154.10 &  \\
7  & 1195.74 & 1192.30 &  \\
8  & 1238.21 & 1234.89 &  \\
9  & 1284.64 & 1281.45 &  \\
10 & 1334.61 & 1331.55 &  \\
11 & 1387.75 & 1384.81 &  \\
12 & 1443.70 & 1440.88 &  \\
13 & 1502.14 & 1499.44 &  \\
14 & 1562.80 & 1560.21 &  \\
15 & 1625.43 & 1622.94 &  \\
16 & 1689.82 & 1687.43 &  \\
17 & 1755.76 & 1753.46 & 1749 $\pm$ 10 \\
18 & 1823.08 & 1820.88 &  \\
19 & 1891.65 & 1889.53 &  \\
20 & 1961.34 & 1959.29 &  \\
21 & 2032.01 & 2030.04 &  \\
22 & 2103.59 & 2101.68 &  \\
23 & 2175.97 & 2174.13 &  \\
24 & 2249.08 & 2247.30 &  \\
25 & 2322.86 & 2321.13 &  \\
26 & 2397.23 & 2395.56 &  \\
27 & 2472.15 & 2470.53 &  \\
28 & 2547.57 & 2546.00 &  \\
29 & 2623.44 & 2621.92 &  \\
30 & 2699.73 & 2698.25 &  \\ \hline
\end{tabular}}
\end{center}

\newpage

\vspace*{10mm}
\begin{center}
Table 30: $M_{17}^{K\eta}(1753-1756)$--Storey.

\vspace{5mm}
\begin{tabular}{|c|c|c|c|r|}\hline
$R(IJ^{P})$ & $ M_R $\, MeV & $ \Gamma_R $\, MeV & Reaction &
Collab.\ \ \\ \hline $K_3^*(\frac{1}{2}3^{-})$ & 1749 $\pm$ 10 &
$193^{\,+\,51}_{\,-\,37}$ & $K^{-}p \rightarrow K^{-}\eta p$ &
LASS 88 \\ \hline
\end{tabular}
\end{center}

\end{document}